\documentclass[twocolumn, showpacs, amsmath,amssymb]{revtex4} 
\usepackage{epsfig, amssymb, graphicx}
\newtheorem{prethm}{{\bf Theorem}}

\newenvironment{thm}{\begin{prethm}{\hspace{-0.5
               em}{\bf.}}}{\end{prethm}}

\newtheorem{prepro}{{\bf Theorem}}%
 \newtheorem{preprop}{{\bf Proposition}}

\newtheorem{precor}{{\bf Corollary}}

\newtheorem{preconj}{{\bf Conjecture}}

\newtheorem{preremark}{{\bf Remark}}

\newtheorem{prelem}{{\bf Lemma}}

\newenvironment{lem}{\begin{prelem}{\hspace{-0.5
               em}{\bf.}}}{\end{prelem}}

\newtheorem{preproof}{{\bf Proof.}}

\newenvironment{proof}[1]{\begin{preproof}{\rm
               #1}\hfill{$\Box$}}{\end{preproof}}

\begin{document}
\preprint{xxx}

\title{Graph States Under the Action of Local Clifford Group in Non-Binary Case}

\author{Mohsen Bahramgiri, Salman Beigi}
\email{m_bahram@mit.edu, salman@math.mit.edu}
\affiliation{Mathematics Department\\ and\\ Computer Science and
Artificial Intelligence
 Laboratory\\
Massachusetts Institute of Technology}
\date{August 20, 2006}

\begin{abstract}
Graph states are well-entangled quantum states that are defined
based on a graph. Of course, if two graphs are isomorphic their
associated states are the same. Also, we know local operations do
not change the entanglement of quantum states. Therefore, graph
states that are either isomorphic or equivalent under the local
Clifford group have the same properties. In this paper, we first
establish a bound on the number of graph states which are neither
isomorphic nor equivalent under the action of local Clifford
group.

Also, we study graph states in non-binary case. We translate the
action of local Clifford group, as well as measurement of Pauli
operators, into transformations on their associated graphs.

Finally, we present an efficient algorithm to verify whether two
graph states, in non-binary case, are locally equivalent or
not.\end{abstract}

\maketitle

\section{Introduction}

{\it Graph states}, forming a universal resourse for quantum
computation based on masurement, have been used in many
applications in quantum information theory as well as quantum
computation, and have been studied extensively. This is due to
the fact that these states not only maintain a rich structure,
but also can be described transparently in different ways.

For example, the notion of {\it entanglement}, the significant
property of quantum systems compared to the classical ones, has
been widely studied for graph states, for instance see
\cite{marc}. On the other hand, the properties of {\it stabilizer
codes}, the most well-studied class of quantum codes, are really
covered by graph states. These states have been studied in this
point of view too, see e.g. \cite{hein}, \cite{marc}.

\subsection{Basic definitions and recent works}

To overview the definitions, recall that Pauli group is the group
generated by Pauli operators, and Clifford group is its
normalizer, generated by Hadamard, CNOT and the phase gates,
\cite{chuang}. A stabilizer state is the common eigenvalue of a
(full rank) abelian subgroup of Pauli group. Since, Clifford
group is the normalizer of Pauli group, any stabilizer state is
sent to another stabilizer state under Clifford operators. On the
other hand, local operators do not change the properties of a
state in the quantum information theory point of view. Therefore,
we say two stabilizer states are equivalent if they can be sent
to each other under some local Clifford operators.

Graph states are a special class of stabilizer states that are
defined based on a graph, and are of interest because of two
reason. First, they can be represented just by a graph which is
both succinct and also, captures all the properties of the state.
Second, it can be proven that any stabilizer state is equivalent
to a graph state under the action of local Clifford group, see
\cite{chuang}. Indeed, it says that in order to study stabilizer
states it is sufficient to study graph states that have a more
simple representation.

As an example, we know that a Clifford operator sends a stabilizer
state to another one. Therefore, translating this action to their
associated graphs, we find a description of local Clifford group
in the language of graphs. This fact is first discovered in
\cite{hein} and \cite{moor1}. It is also shown that Pauli
measurements can be translated in the language of graph theory.

Here, a question arises naturally. Can two graph states be
equivalent under local Clifford group? The answer, by simply
trying some examples, is yes. So as the second question, one may
ask how many graph states can be equivalent to a given one?

There are some recent works to answer these questions, and also
to relate the latter results of graph theory in this direction.
For example, \cite{bouchalg} and \cite{moor2} have presented an
algorithm to recognize whether two given graph states are
equivalent or not.

The other question is that what the number of different graph
states is. Of course, by different graph states we mean states
that are neither isomorphic nor equivalent under the local
Clifford group, and isomorphic states means they are the same by
relabeling their qubits. In \cite{marc} the number of graph
states, which are non-isomorphic and not equivalent, is counted
for $n$ up to 12, where $n$ is the number of qubits. There is no
a known method yet, to count them in general.

\subsection{Our results}

In this paper, we first find a bound on the number of graph
states that are neither isomorphic nor equivalent under the local
Clifford group. It is the first such a bound.

In the second part of this paper we try to answer the same
questions for the non-binary case, i. e. qudits, where $d\neq 2$.
Indeed, Pauli group is well-know for the non-binary quantum
systems, too. Also, Clifford group, same as binary case, is
defined to be the normalizer of the Pauli group, and in
\cite{erik} there is an interesting characterization of this
generalized Clifford group. So that, stabilizer states, and then
graph states, can be defined same as binary case. Although, all
these notions were known (for example see \cite{knill},
\cite{marc} and \cite{ketkar}), but we did not know anything
about the action of local Clifford group in this case.

Here, we first show that any stabilizer state is equivalent to a
graph state under local Clifford group, same as what we had in
binary case. Then, we focus on non-binary graph states and
translate the action of local Clifford group and also Pauli
measurements to some operators on graphs.

After this paper, these operators over graphs have been studied,
and in \cite{us} an efficient algorithm is presented that given
two graph states, determines whether they are equivalent or not.
We present this result here, too.

\section{Non-equivalent graph states}

The notions of stabilizer states and also graphs states (in
binary case) are well-known in the theory and we do not repeat
them here. We just refer the reader to \cite{chuang} and
\cite{cleve} for a full survey on this subject.

It is known that two graph states are equivalent under local
Clifford group if and only if their associated graphs are
equivalent under {\it local complementation} operators (see for
instance \cite{moor1}), by which we mean the following operator.
Suppose that $v$ is a vertex of a graph $G$. To {\it locally
complement} $G$ at $v$, consider all neighbors of $v$ in $G$ and
complement the subgraph of $G$ induced by these vertices to
obtain a new graph, denoted by $G*v$. Two graphs are called {\it
locally equivalent} if one of them is obtained from the other by
a series of local complementations. This notion have been
extensively studied in graph theory, see \cite{bouchet}. Also,
there is a polynomial time algorithm to recognize whether two
graphs are locally equivalent or not, \cite{bouchalg}.

A {\it tree} is a connected graph which contains no cycle. The
following theorem is a significant result on locally equivalent
graphs (see \cite{bouchtree} ).

\begin{thm}\label{tree}{ Any two locally equivalent trees are isomorphic.

 }\end{thm}

By this theorem, if two trees are not isomorphic then they are not
locally equivalent, and then, the number of non-isomorphic trees
on $n$ vertices is a lower bound on the number graphs which are
not locally equivalent. Note that, there are graphs that are not
locally equivalent to a tree, and so, the number of trees is not
a tight bound. However, it is the only known bound of this type.

The following theorem gives us an approximation of the number of
non-isomorphic trees. This bound is proved in \cite{otter}.

\begin{thm}\label{count}{ Let $T_n$ be the number of non-isomorphic
trees on $n$ vertices, then $$
T_n=\frac{\beta^3\alpha^{9/2}}{4\sqrt{\pi}}\cdot
\frac{\alpha^{-n}}{n^{5/2}}+O(\frac{\alpha^{-n}}{n^{7/2}}),
$$ where $\alpha \approx 0.3383219$ and $\beta \approx7.924780$.

 }\end{thm}

We thus obtain the following result, presenting a lower bound for
the number of non-equivalent graph states under the action of
local Clifford group, whose graphs are connected and mutually
non-isomorphic.

\begin{thm}\label{count}{ Let $A_n$ be the number of graph states
which are locally equivalent to some tree. Then for large enough
values of $n$
$$A_n \approx \frac{\beta^3\alpha^{9/2}}{4\sqrt{\pi}}\cdot
\frac{\alpha^{-n}}{n^{5/2}}\approx
0.5349485\frac{\alpha^{-n}}{n^{5/2}},$$

where $\alpha \approx 0.3383219$.

 }\end{thm}

\noindent{\bf Remark.} $A_n$ is a lower bound for $\chi _n$,
where $\chi _n$ is the number of graph states which are not
equivalent under the local Clifford group, and their associate
graphs are connected and non-isomorphic. But, is this bound a good
one?

For a fixed graph $G$, there is an algorithm to count the number
of graphs locally equivalent to $G$ (see \cite{bouchet}). This
number varies from one graph to another. For instance, the number
of graphs locally equivalent to the complete graph over $n$
vertices is $n+1$, but this number for a path of length $n$ is
$O((1+\sqrt{3})^n)$. Hence, it can vary from linear to
exponential, for different graphs.

This observation shows that the problem of finding the exact
number of non-equivalent graphs is relatively a hard one. The
lower bound presented in the remark above is in fact the best
lower bound known so far, which can be compared to the real
values of $\chi _n$ for small $n$'s in Table 1.

{\center

\begin{tabular}{|c|c|c|c|}
  \hline
  n \ & $\frac {log \chi_n}{n} $ & $\frac {log A_n}{n} $ \\ \hline
  5 \ &\ 0.2772  \ \ &\ 0.2197 \ \ \\
  6 \ &\ 0.3996  \ \ &\ 0.2310 \ \ \\
  7 \ &\ 0.4654  \ \ &\ 0.3138 \ \ \\
  8 \ &\ 0.5768  \ \ &\ 0.3612 \ \ \\
  9 \ &\ 0.6763  \ \ &\ 0.4012 \ \ \\
  10\ &\ 0.8049  \ \ &\ 0.4465 \ \ \\
  11\ &\ 0.9643  \ \ &\ 0.4821 \ \ \\
  12\ &\ 1.1714  \ \ &\ 0.5137 \ \ \\ \hline
\end{tabular}

{\bf Table 1.} Values of $\chi_n$ are taken from \cite{marc}.

}

\section{Non-binary Stabilizer Codes}

The theory of non-binary stabilizer codes and non-binary
stabilizer states have been developed, see \cite{knill},
\cite{ketkar}. In this theory, the notion of stabilizer codes and
stabilizer states, as well as graph states are defined based on
Pauli and Clifford groups. In this section, we establish a
description of the action of local Clifford group on graph states
by operations on their associated graphs.

\subsection{Generalized Pauli and Clifford groups}

Through this section, let $p$ be an odd prime number, $\omega=
e^{2\pi i/p}$ and $\mathbb{F}_p$ be the finite field of $p$
elements. Also, we let $\mathbb{C}^p$ to be the Hilbert space of
every particle in the quantum system (qupit), and $\{ \vert
x\rangle ; x\in \mathbb{F}_p\}$ be an orthonormal basis for this
space.

\noindent{\bf Definition.} For $a, b\in \mathbb{F}_p$, define
unitary operators $X(a)$ and $Z(b)$ on $\mathbb{C}_p$ as follows;

$$X(a)\vert x\rangle =\vert x+a\rangle,$$  $$Z(b)\vert x\rangle=\omega
^{bx}\vert x\rangle.$$


The following properties are proved in \cite{ketkar}.

\begin{lem}\label{1}{
 \begin{itemize}
 \item [${\rm (i)}$] $X(a)X(a')=X(a+a'), Z(b)Z(b')=Z(b+b')$ and $X(a)^\dag=X(-a),
 Z(b)^\dag=Z(-b)$.
\item [${\rm (ii)}$] $\{X(a)Z(b) ; a,b\in \mathbb{F}_p\}$
is a basis for the space of linear operators over $\mathbb{C}^p$.
\item [${\rm (iii)}$] $Z(b)X(a)=\omega^{ab}X(a)Z(b)$.

\item [${\rm (iv)}$] $X(a)Z(b)$ and $X(a')Z(b')$ commute iff $ab'-ba'=0$.
\end{itemize}

}\end{lem} \vspace{1mm}

Using these properties, we can define the {\it generalized Pauli
group}, generated by these operators.
$$\mathcal{G}=\{\omega^c X(a)Z(b) ; a,b,c\in \mathbb{F}_p\}.$$ Also, the
{\it Pauli group over $n$ qupits} is the $n$-fold tensor product
of $\mathcal{G}$
$$\mathcal{G}_n=\{\omega^c X(\bold {a})Z(\bold{b}) ; \bold{a},\bold{b}\in
{\mathbb{F}_p ^n}, c\in \mathbb{F}_p\},$$ where
$X(\bold{a})=X(\bold{a}_1) \otimes X(\bold{a}_2) \otimes \cdots
\otimes X(\bold{a}_n)$ and $Z(\bold{b})=Z(\bold{b}_1) \otimes
Z(\bold{b}_2) \otimes \cdots \otimes
 Z(\bold{b}_n)$.

By lemma \ref{1}, part ${\rm (iv)}$, one can easily check that two
elements $\omega^c X(\bold {a})Z(\bold{b})$ and $\omega^{c'}
X(\bold {a'})Z(\bold{b'})$ commute if and only if $\bold{a}\cdot
\bold{b'}-\bold{b}\cdot \bold{a'}=0$ (dot product is the usual
inner product on ${\mathbb{F}_p ^n}$, i.e.
$\displaystyle\sum_{i=1}^{n}
\bold{a}_i\bold{b'}_i-\bold{a'}_i\bold{b}_i=0$ ).

\vspace{5mm}

\noindent{\bf Definition.} {\it Generalized Clifford group}
$\mathcal{C}_n$ is the normalizer of $\mathcal{G}_n$. Also,
generalized {\it local Clifford group} is the $n$-fold tensor
product of Clifford group of order one,
($\mathcal{C}_1=\mathcal{C}$).

\vspace{3mm}

In the binary case we know that the Clifford group is generated
by Hadamard, CNOT and the phase gates. For the general case, in
order to somehow characterize $\mathcal{C}$, suppose $h\in
\mathcal{C}$. Then by definition, $hX(1)h^\dag$ and $hZ(1)h^\dag$
are in $\mathcal{G}$. Let $hX(1)h^\dag=\omega^c X(a)Z(b)$ and
$hZ(1)h^\dag=\omega^{c'} X(a')Z(b')$. Since $Z(1)X(1)=\omega
X(1)Z(1)$, by lemma \ref{1}, we have $ab'-ba'=1$. The following
theorem states that, this is the only condition on $h$ to make it
an element of $\mathcal{C}$. See \cite{erik} for the proof.

\begin{thm}\label{clif}{ For any $a,b,c,a',b'$ and $c'$ in
$\mathbb{F}_p$, such that $ab'-ba'=1$, there exists $h\in
\mathcal{C}$ such that $hX(1)h^\dag=\omega^c X(a)Z(b)$ and
$hZ(1)h^\dag=\omega^{c'} X(a')Z(b')$.

 }\end{thm}

\subsection{Stabilizer states}

Before introducing the notion of stabilizer codes, we should first
study some properties of eigenvalues and eigenspaces of Pauli
operators.

\begin{lem}\label{2}{
 Let $ g=\omega^c X(\bold{a})Z(\bold{b}) \in \mathcal{C}_n $. Then for every
 positive integer $k$, $$g^k=\omega^{(kc+{k \choose 2}\bold{a.b})}
 X(k\bold{a})Z(k\bold{b}),$$ where ${k \choose 2}$ is the binomial
 coefficient.

}\end{lem}

\begin{proof}{
By induction on $k$ we prove the lemma. There is nothing to prove
for $k=1$, and for $k+1$ we have
  $$g^{k+1}=\big(\omega^cX(\bold{a})Z(\bold{b})\big)\big(\omega^{(kc+{k \choose
2}\bold{a.b})}
 X(k\bold{a})Z(k\bold{b})\big)$$
 $$=\omega^{((k+1)c+{k\choose
2}\bold{a.b})}X(\bold{a})\big(Z(\bold{b})X(k\bold{a})\big)Z(k\bold{b})$$
 $$=\omega^{((k+1)c+{k\choose
2}\bold{a.b})}X(\bold{a})\big(\omega^{k\bold{a.b}}X(k\bold{a})Z(\bold{b})\big)Z(k\bold{b})$$
 $$=\omega^{((k+1)c+{k+1 \choose 2}\bold{a.b})}
 X((k+1)\bold{a})Z((k+1)\bold{b}).$$
}\end{proof}

Since, by definitions, $X(p\,\bold{a})=Id$, $Z(p\,\bold{b})=Id$
and $\omega^p=1$, one obtains that $g^p=Id$ for any $g\in
\mathcal{G}_n$. Notice that, we are now using the fact that $p$ is
odd. In fact, for an odd number $m$, $m$ divides ${m \choose 2}$.
But it is not true for any even number. It is why we should add
multiplicity of $i=\sqrt{-1}$ in the definition of Pauli group in
the binary case. This observation shows that eigenvalues of the
elements of $\mathcal{G}_n$ are $p$-th roots of the unity.

\begin{lem}\label{3}{
 Suppose that $g=\omega^c X(\bold{a})Z(\bold{b})\in \mathcal{G}_n$ is not a scalar
 multiple of the identity. Let $$P_j=\frac{1}{p}(Id+\omega ^{-j}g+\omega ^{-2j}g + \cdots +\omega
 ^{-(p-1)j}g)$$ for $j=0,1,\dots (p-1)$. Then $P_j$ is the projection on $\omega ^j$-eigenspace
 of g. Also, all $P_j$'s have same ranks.

}\end{lem}

\begin{proof}{
Since $g^p=Id$, clearly $P_j^2=P_j$ and $gP_j=\omega ^j P_j$.
Therefore, it is sufficient to prove that all $P_j$'s have equal
ranks. Since, $g$ is not a scaler multiple of the identity, at
least one of $\bold{a}_i$'s or $\bold{b}_i$'s is non-zero. For
instance, suppose that  $\bold{a}_i\neq 0$ (the other case is
similar). One can simply check that
$$Z_i(k)P_jZ_i(-k)=P_{j-k\bold{a}_i}$$ holds for every $k$.
Finally, since $\bold{a}_i$ is non-zero, all $P_j$'s are
conjugate and thus have the same rank.

 }\end{proof}

We can now define the stabilizer code as the common eigenspace
for eigenvalue one, of a subgroup of $\mathcal{G}_n$. It is easy
to check that for a subgroup $S$, similar to the binary case, this
common eigenspace is non-trivial provided that $S$ is abelian and
does not contain any scalar multiple of identity, except $Id$
itself, see \cite{ketkar}. We call such a subgroup a {\it valid}
one.

In order to define graph states, we should investigate the
properties of stabilizer codes more precisely. Consider a valid
subgroup $S$ of $\mathcal{G}_n$. Let $S=\langle g_1,\dots
g_k\rangle$ be a minimal set of generators for $S$, so that no
subset of $\{g_1,\dots g_k\}$ generates $S$. Suppose that
$$g_i=\omega^{c_i} X(\bold{a^i})Z(\bold{b^i}).$$ Since, the only
scalar multiple of the identity in $S$ is itself, for any
$\bold{a,b} \in \mathbb{F}_p^n$, at most one element of $S$ is of
the form $\omega^c X(\bold{a})Z(\bold{b})$. It means that, without
any ambiguity, in order to represent the elements of $S$, we can
drop the scaler coefficient $\omega^c$, and for any $\omega^c
X(\bold{a})Z(\bold{b}), \omega^{c'} X(\bold{a'})Z(\bold{b'})\in
S$, we can write $$ (X(\bold{a})Z(\bold{b}))
(X(\bold{a'})Z(\bold{b'}))\equiv X(\bold{a+a'})Z(\bold{b+b'}),$$
as an equality in $S$. Therefore, by this notation, set of vectors
$(\bold{a}, \bold{b})$ where $X(\bold{a})Z(\bold{b})$ is in $S$
consist a vector subspace of $\mathbb{F}_p^{2n}$. In fact, since
$S$ is generated by $g_1, \dots g_k$ this vector subspace is
generated by vectors $(\bold{a^i,b^i})$, $i=1,\dots k$. So, it
would be helpful to define the matrix

$$A=\begin{pmatrix}
   &  \bold{a^1}&  & \vline &  & \bold{b^1} &   \\
   &  \bold{a^2}&  & \vline &  & \bold{b^2} &   \\
   &  \vdots    &  & \vline &  & \vdots     &   \\
   &  \bold{a^k} & & \vline &  & \bold{a^k} &
\end{pmatrix}$$
as a {\it generator matrix} for $S$.

In general, a matrix is a generator matrix of $S$ if for any
$(\bold{a},\bold{b})$ a row of matrix, there exists $c$ such that
$\omega^c X(\bold{a})Z(\bold{b})$ is in $S$, and also, these
operators for different rows consist a minimal set of generators.


\begin{lem}\label{4}{ \begin{itemize}
\item [${\rm (i)}$] Using the above notation, $S=\{xA ;\ x\in
\mathbb{F}_p^k\}$.
\item [${\rm (ii)}$] The matrix $UA$ is also a generator matrix of $S$, for
any invertible $k\times k$ matrix $U$. Moreover, any generator
matrix of $S$ is of this form.
\item [${\rm (iii)}$] Rows of $A$ are orthogonal with respect to
the inner product defined as
$\langle(\bold{a,b}),(\bold{a',b'})\rangle=\bold{a.b'-b.a'}$.
\item [${\rm (iv)}$] $rank(A)=k$.

\end{itemize}
}\end{lem}

\begin{proof}{ ${\rm (i)}$ follows from the definition, and ${\rm
(ii)}$ comes from the fact that any element of $S$ should be of
the form $xA$, (part ${\rm (i)}$). ${\rm (iii)}$ is true because
$S$ is abelian, and ${\rm (iv)}$ follows from ${\rm (iii)}$.

}\end{proof}

\begin{lem}\label{5}{
Suppose that $P_{ij}$ is the projection over the $\omega
^j$-eigenspace of $g_i$. Then $P_{10}P_{20}\cdots P_{k0}$ is the
projection over the code space. Also, all $P_{1j_1}P_{2j_2}\cdots
P_{kj_k}$'s have the same rank.

}\end{lem}

\begin{proof}{
All $g_i$'s commute, and therefore by lemma \ref{3}, all
$P_{ij}$'s commute as well. So, the first argument follows
immediately. Since $(\bold{a^i, b^i})$'s are independent by lemma
\ref{4}, there exists $(\bold{e^i, f^i})$ such that
$h_i=X(\bold{e^i})Z(\bold{f^i})$ commutes with each $g_l$, where
$l \neq i$, and also, $h_i g_i h_i ^\dagger=\omega ^{r_i}g_i$ for
non-zero $r_i$. In other words, $(\bold{e^i, f^i})$ is orthogonal
to all $(\bold{a^l, b^l})$'s but $(\bold{a^i, b^i})$. Now let
$h=h_1h_2\cdots h_k$, we have $hP_{1j_1}P_{2j_2}\cdots
P_{kj_k}h^\dagger = P_{1(j_1-r_1)}P_{2(j_2-r_2)}\cdots
P_{k(j_k-r_k)}$. Since $r_i$'s are arbitrary, all
$P_{1j_1}P_{2j_2}\cdots P_{kj_k}$'s are conjugate, and therefore
have the same rank.

}\end{proof}

Using this lemma, we conclude that $rank (P_{10}P_{20}\cdots
P_{k0})=p^{n-k}$. It means that, a stabilizer group with $k$
generators corresponds to a stabilizer code of dimension $n-k$.
Therefore, if we have $n$ independent generators, we get a one
dimensional code space, i.e. a {\it stabilizer state}.

\subsection{Graph states}

Next, we consider the action of the local Clifford group on
stabilizer spaces. If $h=h_1h_2\cdots h_n \in \mathcal{C}^{\otimes
n}$ is an element of the local Clifford group, then $hSh^\dagger$
is also an abelian subgroup of $\mathcal{G}_n$, and the only
scalar multiple of the identity in $hSh^\dagger$ is $Id$ itself.
In fact, if $S$ is the stabilizer group of the state $\vert \phi
\rangle$, then $hSh^\dagger$ is the stabilizer group of $h\vert
\phi \rangle$.

Since $hSh^\dagger$ is a stabilizer group, it has a generator
matrix, and our goal is to compute the generator matrix of this
group in terms of $A$, the generator matrix of $S$. Indeed, $g_1,
\dots g_n $ are generators of $S$. So that, $hg_1h^\dagger,\dots
hg_nh^\dagger$ are generators of $hSh^\dagger$. Thus, suppose
$$h_iX(1)h{_i}^{\dagger}=\omega^{d_i} X(e_i)Z(f_i),$$
$$h_iZ(1)h_i^{\dagger} =\omega^{d'_i} X(e'_i)Z(f'_i).$$ By theorem
\ref{clif}, we have $e_i f'_i - f_i e'_i=1$, and by the above
observation it is a simple calculation to check that the generator
matrix of $hSh^\dagger$ is the matrix $AY$, where
$$Y=\begin{pmatrix}
  E & F \\
  E' & F'
\end{pmatrix},$$ and
$$E=diag(e_1,\cdots, e_n), F=diag(f_1,\cdots, f_n),$$
$$E'=diag(e'_1,\cdots, e'_n), F'=diag(f'_1,\cdots, f'_n).$$

\begin{lem}\label{6}{
Two stabilizer states with generator matrices $A, B$ are
equivalent under the action of the local Clifford group, if and
only if there exist invertible matrices $U$ and $Y$, such that
$$Y=\begin{pmatrix}
  E & F \\
  E' & F'
\end{pmatrix},$$ where $$E=diag(e_1,\cdots, e_n), F=diag(f_1,\cdots,
f_n),$$ $$E'=diag(e'_1,\cdots, e'_n), F'=diag(f'_1,\cdots,
f'_n),$$ and $e_i f'_i - f_i e'_i=1$, for any $i$, and also,
$B=UAY$ holds as well.

}\end{lem}

\begin{proof}{ The proof follows from the above discussion
together with lemma \ref{4}.

}\end{proof}

This lemma can be restated in the following way

\noindent\textbf{Lemma 6$'$.}{\it \ Two stabilizer states with
generator matrices $A, B$ are equivalent under the action of the
local Clifford group, if and only if there exists an invertible
matrix $Y$, such that
$$Y=\begin{pmatrix}
  E & F \\
  E' & F'
\end{pmatrix},$$ where $$E=diag(e_1,\cdots, e_n), F=diag(f_1,\cdots,
f_n),$$ $$E'=diag(e'_1,\cdots, e'_n), F'=diag(f'_1,\cdots,
f'_n),$$ and $e_i f'_i - f_i e'_i=1$, for any $i$, and also, all
rows of $B$ are orthogonal to rows of $AY$. }

\begin{proof}{ By lemma \ref{4}, rows of $B$ are orthogonal to
each other, and $rank(B)=n$. Therefore, rows of $B$ consist a
vector subspace of dimension $n$, in the space of dimension $2n$,
and is orthogonal to itself. So that, any vector orthogonal to
rows of $B$ is in this subspace.

Now suppose that, rows of $AY$ are orthogonal to rows of $B$.
Hence, rows of $AY$ are in the subspace generated by $B$, and
since $rank(AY)=n$, there exists an invertible matrix $U$ such
that $B=UAY.$ The other direction follows from lemma \ref{6}.

}\end{proof}

 We can now define the notion of
{\it graph state}, and its associated labeled graph.

\noindent{\bf Definition.} A {\it graph state} is the stabilizer
state of a group with a generator matrix of the form
$\begin{pmatrix}
  Id_n & \mid & M
\end{pmatrix}$, where $M$ is a symmetric matrix with zero
diagonal. Note that, this matrix has rank $n$, because of
identity matrix in the first block. Also, all of whose rows are
orthogonal since $M$ is symmetric. Hence, $\begin{pmatrix}
  Id_n & \mid & M
\end{pmatrix}$ is really a generator matrix of a stabilizer
group.

We assign to such a graph state a labeled graph over $n$ vertices
$\{1, \dots n\}$ , with labels coming from the matrix $M$, i.e.,
with label $M_{ij}$ for the edge $\{ij\}$.

\begin{lem}\label{7}{
Every stabilizer state is equivalent to a graph state with
respect to the local Clifford group.

}\end{lem}

\begin{proof}{
Consider a stabilizer state with generating matrix $A$. By lemma
\ref{4}, $A$ is full-rank and the rows of $A$ are orthogonal.
Moreover, since by lemma \ref{6} we can apply a linear
transformation of determinant one on any pair of columns $i$ and
$(n+i)$, we may assume that the first $n\times n$ block of $A$ is
invertible. Then, we apply an invertible matrix $U$ such that the
first block of $UA$ is identity. Using the fact that the rows of
$UA$ are orthogonal, we conclude that the second block of $UA$ is
symmetric. Notice that, even though this symmetric matrix may have
non-zero diagonal entries, but by applying the determinant one
linear operations on pair of columns, we can end up with a matrix
with identity in the first block, and a symmetric matrix with
zero diagonal in the second block.

}\end{proof}

\subsection{Description of local Clifford group in terms of operations over graphs}

In this section, similar to the binary case in \cite{moor1}, we
want to describe the action of local Clifford group in terms of
operations on graphs. First, we should define the operators.

\vspace{3mm}

\begin{figure}[htp]
\begin{center}
\includegraphics[width=3in]{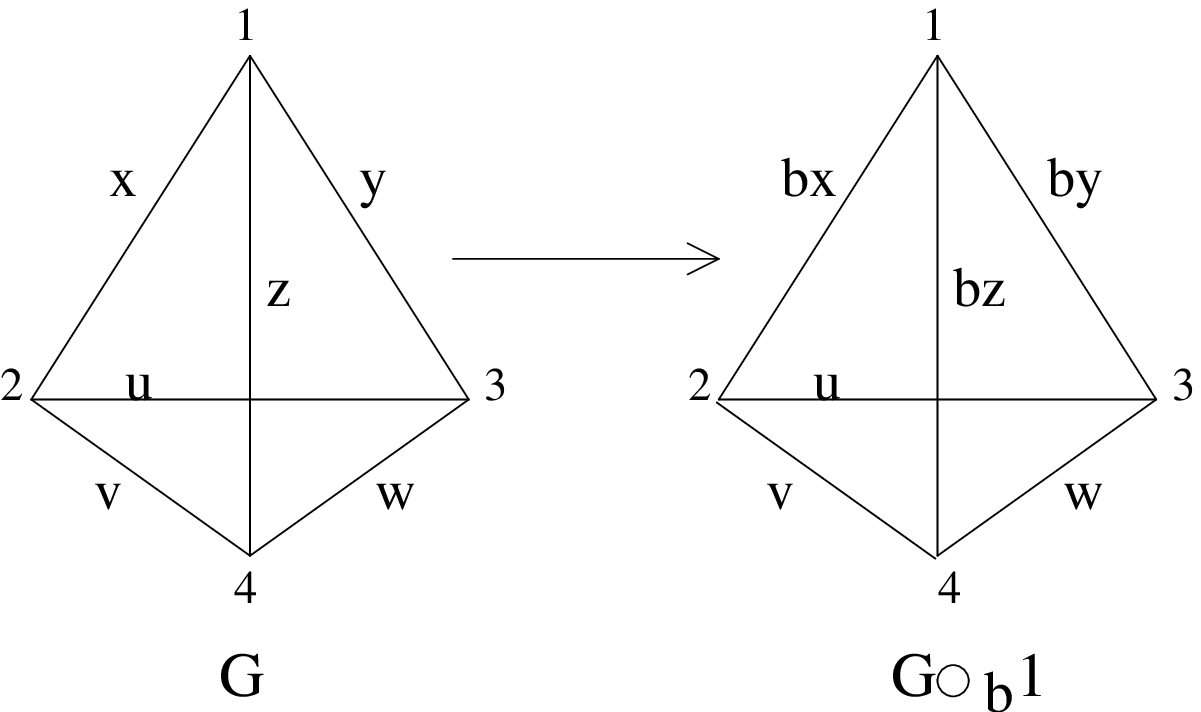}
\caption{ Graph $G$ after applying operator $\circ_b1$}
\label{fig:1}
\end{center}
\end{figure}

\noindent{\bf Definition.} Let $G$ be a labeled graph on vertex
set $\{1,2, \dots n \}$, such that the label of the edge
$\{i,j\}$ is the $ij$-th entry of matrix $M$, where $M$ is a
symmetric matrix over $\mathbb{F}_p$ with zero diagonal. For
every vertex $v$, and $0\neq b\in \mathbb{F}_p$, define the
operator $\circ_b v$ on the graph as follows; $G \circ_b v$ is
the graph on the same vertex set, with label matrix $I(v,b)M
I(b,a)$, where $I(v,b)=diag(1,1, \dots,b, \dots,1)$, $b$ being on
the $v$-th entry. See figure \ref{fig:1}.

\begin{figure}[htp]
\begin{center}
\includegraphics[width=3in]{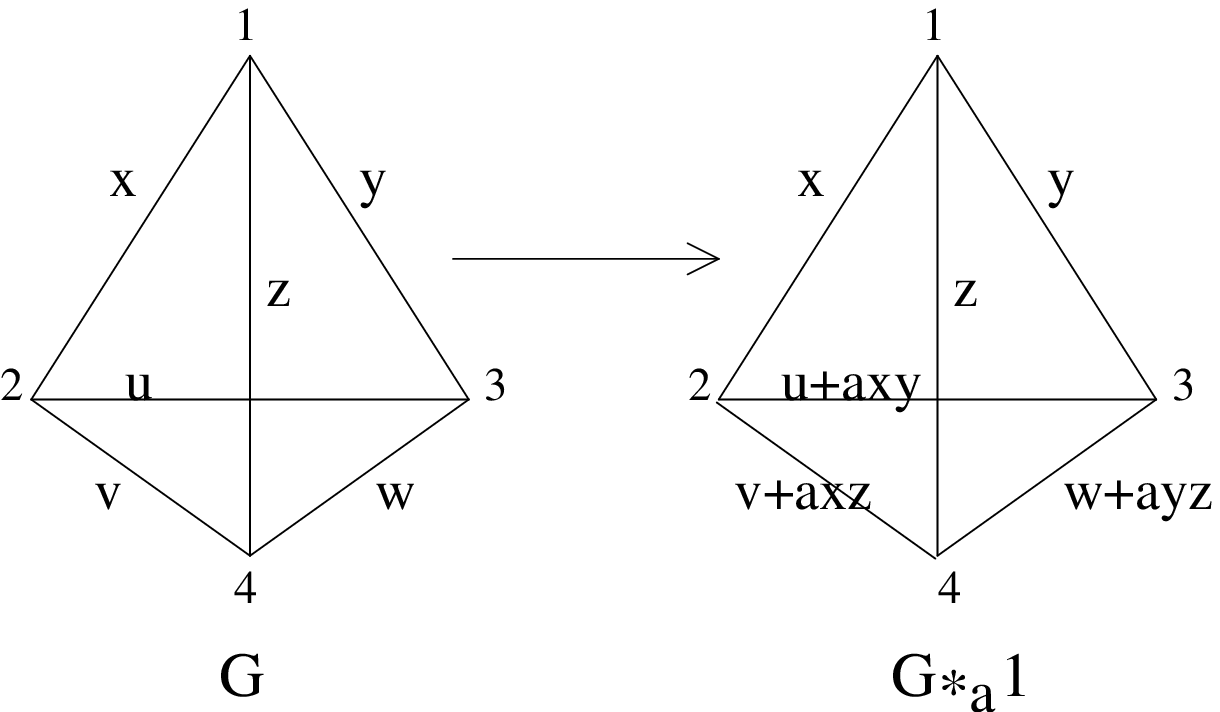}
\caption{Graph $G$ after applying operator $*_a1$} \label{fig:2}
\end{center}
\end{figure}

Also, for every vertex $w$, and $a\in \mathbb{F}_p$ define the
operator $*_a w$ on the graph as follows; $G *_a w$ is the graph
on the same vertex set, with label matrix $M'$, where
$M'_{jk}=M_{jk}+aM_{vj}M_{vk}$ for $j\neq k$, and $M'_{jj}=0$ for
all $j$. See figure \ref{fig:2}.

Now, the main theorem;

\begin{thm}\label{nongraph}{ Two graph states $G$ and $H$ with label matrices
$M$ and $N$ over $\mathbb{F}_p$, are equivalent under local
Clifford group if and only if there exists a sequence of $*$ and
$\circ$ operators acting on one of them to obtain the other.

 }\end{thm}

\begin{proof}{
Let $A=\begin{pmatrix}
  Id_n & \mid & M
\end{pmatrix}$ and $B=\begin{pmatrix}
  Id_n & \mid & N
\end{pmatrix}$ be the generator matrices of these two graph
states. If these two states are equivalent, by lemma \ref{6},
there exist matrices $U$ and $Y=\begin{pmatrix}
  E & F \\
  E' & F'
\end{pmatrix}$ satisfying the conditions mentioned in lemma \ref{6}, such that $B=UAY$.


For every $i$, let $$Y_i=\begin{pmatrix}
  E_{i} & F_{i} \\
  E'_{i} & F'_{i}
\end{pmatrix},$$ where $$E_i=diag(1,\cdots ,1,e_i,\cdots,1, 1),$$  $$F_i=diag(0, \cdots,
0,f_i,0,\cdots, 0),$$ $$E'_i=diag(0,\cdots,0, e'_i,0,\cdots ,0),$$
$$F'_i=diag(1,\cdots,1, f'_i,1,\cdots, 1).$$ Then $Y_i$'s commute
mutually, and $Y=Y_1Y_2\cdots Y_n$. We call $Y_i$ {\it trivial} if
$E_i=Id_n$ and $E_i'=0$.

We prove the theorem by induction on the number of non-trivial
matrices $Y_i$'s. If all $Y_i$'s are trivial, then
$AY=\begin{pmatrix}
  Id_n & \mid &D
\end{pmatrix}$, for some matrix $D$. Therefore $U=Id_n$, as well as
$Y=Id_{2n}$ and $A=B$. Thus, suppose that at least one of the
$Y_i$'s is non-trivial.

We consider two cases;

\noindent {\bf Case (i).} $e_{i_0}\neq 0$ for some $i_0$, where
$Y_{i_0}$ is non-trivial. Let $AY_{i_0}=\begin{pmatrix}
  V & \mid & D
\end{pmatrix}$. Therefore,

$$V=\begin{pmatrix}
  1      & 0  & \cdots & e'_{i_0}M_{1i_0}& \cdots & 0      &   \\
  0      & 1  &        &  \vdots         &        & 0      &   \\
  \vdots &    & \ddots &  \vdots         &        & \vdots &   \\
  \vdots &    &        & e_{i_0}         &        & \vdots &   \\
  \vdots &    &        &  \vdots         & \ddots & 0      &   \\
  0      & 0  & \cdots & e'_{i_0}M_{ni_0}& \cdots & 1      &
\end{pmatrix}.$$

In order to invert $V$, one has to multiply the $i_0$-th row by
$e^{-1}_{i_0}$, and then multiply it to $-e'_{i_0}M_{ji_0}$ and
finally add it to $j$-th row, for any $j\neq i_0$.

Therefore, $V^{-1}AY_{i_0}=\begin{pmatrix}
  Id_n & \mid & V^{-1}D
\end{pmatrix}$. Also, the $jk$ entry of $V^{-1}D$, for $j\neq k$ and both unequal to $i_0$, is
$$(V^{-1}D)_{jk}=M_{jk}-e'_{i_0}e^{-1}_{i_0}M_{ij}M_{ik},$$ and the $i_0j$ entry
is $$(V^{-1}D)_{i_0j}=e^{-1}_{i_0}M_{ij}.$$ $V^{-1}D$ may have
non-zero entries on its diagonal. But there exists a trivial
matrix $Y'$, such that $V^{-1}AY_{i_0}Y'$ is equal to
$V^{-1}AY_{i_0}$, except on the entries of the diagonal of the
second block, which are all zero for the matrix
$V^{-1}AY_{i_0}Y'$. Therefore $V^{-1}AY_{i_0}Y'$ is the generator
matrix of a graph state, and by the above equalities, this graph
state is
$$G *_{i_0} (-e^{-1}_{i_0}e'_{i_0}) \circ_{i_0}
e^{-1}_{i_0}$$ On the other hand, we have
$$B=UAY=UV(V^{-1}AY_{i_0}Y')(Y'^{-1}Y''),$$ where $Y''$ is equal to the
multiplication of all $Y_j$'s except $Y_{i_0}$. We now observe
that $V^{-1}AY_{i_0}Y'$ is a graph obtained from $G$, via
operations $*$ and $\circ$. Also the number of non-trivial terms
in $Y'^{-1}Y''$ is less than this number in $Y$, and therefore by
induction, the claim is proved.

\noindent{\bf Case (ii).} $e_{i}= 0$ for all $i$'s that $Y_{i}$ is
non-trivial. Suppose that $Y_{i_0}$ is non-trivial. If for every
non-trivial $Y_j$, $M_{i_{0}j}= 0$, then the $i_0$-th row of the
first block of $AY$ is zero and hence, it would not be invertible
and the first block of $UAY$ can not be the identity. Thus, there
exists an $i_1$, such that $Y_{i_1}$ is non-trivial and
$M_{i_{0}i_1}\neq 0$. Then the first block of the matrix
$AY_{i_0}Y_{i_1}$ is
$$V=\begin{pmatrix}
  1      &\cdots & e'_{i_0}M_{1i_0}   & e'_{i_1}M_{1i_1}  & \cdots & 0      &   \\
  0      &\ddots & \vdots             &  \vdots           &        & 0      &   \\
  \vdots &       & 0                  & e'_{i_1}M_{i_0i_1}&        & \vdots &   \\
  \vdots &       & e'_{i_0}M_{i_1i_0} &    0              &        & \vdots &   \\
  0      &       & \vdots             &  \vdots           & \ddots & 0      &   \\
  0      &\cdots & e'_{i_0}M_{ni_0}   & e'_{i_1}M_{ni_1}  & \cdots & 1      &
\end{pmatrix}$$

In order to invert $V$, one has to multiply the $i_0$-th and the
$i_1$-st rows to $(e'_{i_1}M_{i_0i_1})^{-1}$ and
$(e'_{i_0}M_{i_1i_0})^{-1}$, respectively, and then multiply the
$i_1$-st row to $-e'_{i_0}M_{ji_0}$ and add it to the $j$-th row,
for any $j$. And the same process for the $i_0$-th row. Notice
that $e_{i_0}f'_{i_0}-e'_{i_0}f_{i_0}=1$ and $e_{i_0}=0$, so
$e'_{i_0}f_{i_0}=-1$ and consequently, $e'_{i_0}$ is non-zero.
Also the same is true for $e'_{i_1}$. After all, switch the rows
$i_0$ and $i_1$. By this process we get a matrix with the
identity in the first block as well as a symmetric matrix on the
second block. The non-zero elements of the diagonal of this block
can be fixed by multiplying by an appropriate trivial matrix $Y'$,
same as previous case. So, we get
$$A'=V_{-1}AY_{i_0}Y_{i_1}Y'=\begin{pmatrix}
  Id_n & \mid & M'
\end{pmatrix},$$ where $M'$ is the matrix of a graph $G'$ such that
$$M'_{jk}=M_{jk}-M_{i_0i_1}^{-1}M_{i_0j}M_{i_1k}-M_{i_0i_1}^{-1}M_{i_1j}M_{i_0k}.$$
Now, one can see that

$$G'= G \circ_{(-M^{-1}_{i_0i_1})} i_0 *_{1} i_0 *_{(-1)} i_1 *_1 i_0 \circ_{(e'^{-1}_{i_0}M^{-1}_{i_0i_1})} i_0
\circ_{(e'^{-1}_{i_1})} i_1.$$

We have $B=UV^{-1}A'Y'^{-1}Y''$, where $Y''$ is equal to the
multiplication of all $Y_j$'s, except $Y_{i_0}$ and $Y_{i_1}$.
Also the number of non-trivial terms in $Y'^{-1}Y'$ is strictly
less than this number in $Y$, and therefore, by induction, the
claim is proved.

The other direction of the theorem is an immediate consequence of
the first direction.

}\end{proof}

\subsection{Local measurement of Pauli operators}

Theorem \ref{nongraph} tells us that the action of the local
Clifford group can be translated into operations on graphs. We do
the same process, for local measurement of Pauli operators.

Suppose that the stabilizer group of a state is generated by
$g_1, g_2, \cdots ,g_n$ and we measure $h\in \mathcal{G}_n$.
Assume that $g_i^{-1}hg_i=\omega^{c_i}h$. If $c_i$'s are all zero
then $h$ commutes with all of $g_i$'s, and therefore the outcome
of the measurement is the state itself, with the unchanged
stabilizer group. Otherwise, there exists at least one non-zero
$c_i$. By changing the set of generators for stabilizer group
(using lemma \ref{4}), we can assume that $c_1$ is non-zero, and
$c_i=0$, $i=2,\cdots, n$. Therefore $\omega^{-c}h, g_2,\cdots,
g_n$ is a set of generators for the state after the measurement,
in which $c$ is the outcome of measurement. We use this idea to
translate the local Pauli measurements to operations on graphs. In
order to do so, we need the following definition.

\noindent{\bf Definition.} Suppose that $G$ is a labeled graph on
$\mathbb{F}_p$. If $v$ is a vertex of $G$, define $d(v)G$ to be a
graph on the same vertex set, but all neighborhood edges of $v$
have zero labels.

\begin{thm}\label{measx}{ Suppose we have a graph state with label
matrix $M$ and we measure the operator $X_i(a)$, that is $X(a)$ on
the $i$-th qupit. Then if $M_{ij}=0$ for all $j=1,\dots, n$ then
the state remains unchanged, and if $M_{ij}\neq 0$ for some $j$,
then the state after the measurement is equivalent to

$$d(i) \big{(} G \circ_{(M_{ij}^{-2})} i  \circ_{(-M_{ij})} j \circ_{(M_{ij}^{-2})} i \big{)}.$$

 }\end{thm}

\begin{proof}{
If $M_{ij}=0$, then $X_i(a)$ commutes with the stabilizer group,
and therefore, the state remains unchanged after the measurement.
Thus, let us suppose that $M_{ij}\neq 0$ for some $j$, and let
$\alpha=M_{ij}$. Now $U\begin{pmatrix}
  Id_n & \mid & M
\end{pmatrix}$ is also a generator matrix for the graph state (lemma \ref{4}), where $$U=\begin{pmatrix}
  1      & 0  & \cdots & -\alpha^{-1}M_{1i}& \cdots & 0      &   \\
  0      & 1  &        & -\alpha^{-1}M_{2i}&        & 0      &   \\
  \vdots &    & \ddots &  \vdots           &        & \vdots &   \\
  \vdots &    &        & 1                 &        & \vdots &   \\
  \vdots &    &        &  \vdots           & \ddots & 0      &   \\
  0      & 0  & \cdots & -\alpha^{-1}M_{ni}& \cdots & 1      &
\end{pmatrix},$$
and non-zero off-diagonal entries are on $j$-th column. We observe
that $X_i(a)$ commutes with all of the rows of $U\begin{pmatrix}
  Id_n & \mid & M
\end{pmatrix}$, except the $j$-th one. Hence, if we replace the $j$-th
row by $$(0,\cdots,0,a,0,\cdots,0 \mid 0,\cdots,0),$$ we obtain
the generator matrix for the new state. On the other hand, by
lemma \ref{7}, every stabilizer state is equivalent to a graph
state under the local Clifford group, and if we apply this process
to the new generator matrix we end up with the generator matrix
$\begin{pmatrix}
  Id_n & \mid & M'
\end{pmatrix}$, where $M'$ is the matrix of the the graph
$$d(i) \big{(} G \circ_{(M_{ij}^{-2})} i  \circ_{(-M_{ij})} j \circ_{(M_{ij}^{-2})} i \big{)}.$$

}\end{proof}

The changes after measuring the operator $X_i(a)Z_i(b)$ on a
graph state is presented in the following theorem.

\begin{thm}\label{measxz}{ Consider a graph state with label
matrix $M$, and suppose that one measures the operator
$X_i(a)Z_i(b)$ on the $i$-th qupit, where $a, b$ are non-zero.
Then the state after measurement is equivalent to $d(i) \big{(} G
\circ_{(ab^{-1})} i \big{)}$.

 }\end{thm}

\begin{proof}{
First, suppose that $M_{ij}=0$ for every $j$. In this case
$X_i(a)Z_i(b)$ commutes with all of the generators, except the
$i$-th one. So, if we replace it by
$$(0,\cdots,0,a,0,\cdots,0 \mid 0,\cdots,0,b,0,\cdots,0),$$
where $a$ and $b$ both locate on the $i$-th entries, then, using
the local Clifford group, we obtain the same generator matrix and
the same graph. Notice that in this case, $d(i) \big{(} G
\circ_{(ab^{-1})} i \big{)}$ is same as $G$.

Therefore, assume that $M_{ij}\neq 0$ for some $j$, and let
$\alpha=M_{ij}$. Clearly, $(U+\alpha
a^{-1}b\delta_{ij})\begin{pmatrix}
  Id_n & \mid & M
\end{pmatrix}$ is also a generator matrix for the stabilizer
group, where $$U=\begin{pmatrix}
  1      & 0  & \cdots & -\alpha^{-1}M_{1i}& \cdots & 0      &   \\
  0      & 1  &        & -\alpha^{-1}M_{2i}&        & 0      &   \\
  \vdots &    & \ddots &  \vdots           &        & \vdots &   \\
  \vdots &    &        & 1                 &        & \vdots &   \\
  \vdots &    &        &  \vdots           & \ddots & 0      &   \\
  0      & 0  & \cdots & -\alpha^{-1}M_{ni}& \cdots & 1      &
\end{pmatrix},$$
and again, non-zero off-diagonal entries are on $j$-th column. It
is easy to check that all of the rows of $(U+\alpha
a^{-1}b\delta_{ij})\begin{pmatrix}
  Id_n & \mid & M
\end{pmatrix}$, except the $j$-th one, commute with $X_i(a)Z_i(b)$. Hence,
if we replace it by the row-representation of $X_i(a)Z_i(b)$ we
get to a generator for the new state. Applying the algorithm to
get a graph state from this state (lemma \ref{7}), we obtain that
the new graph is $d(i) \big{(} G \circ_{(ab^{-1})} i \big{)}$.

}\end{proof}

And finally, measuring the operator $Z_i(b)$ has some affects on
the graph states, described below in details.

\begin{thm}\label{measz}{ Suppose that $G$ is a graph state with label
matrix $M$, and we measure the operator $Z_i(b)$ on the $i$-th
qupit, where $b$ is non-zero. Then the state after this
measurement is equivalent to $d(i)G$.

 }\end{thm}

\begin{proof} { Consider the generating matrix $\begin{pmatrix} Id_n & \mid & M
\end{pmatrix}$ of the graph state. We know that all of the rows of
this matrix, except the $i$-th one, are orthogonal to the
row-representation of $Z_i(b)$, which is $(0,\dots,0 \mid
0,\dots,b,0,\dots 0)$. Therefore, after the measurement, the
stabilizer group is in fact the group generated by the rows of
$\begin{pmatrix} Id_n & \mid & M
\end{pmatrix}$, except the $i$-th one, and $Z_i(b)$. Therefore, the
stabilizer state, after deleting the $i$-th qupit, is $d(i)G.$

}\end{proof}

\section{Efficient algorithm to recognize equivalency of graph states}


Let $G$ and $H$ be two connected graphs, with label
matrices $M$ and $N$, and assume that these two graphs are equivalent under the action of local
Clifford group. Notice that by operations
$*$ and $\circ$, a connected graph remains connected. Therefore, by lemma 6$'$, There exists a matrix
$Y$ such that
$$Y=\begin{pmatrix}
  E & F \\
  E' & F'
\end{pmatrix},$$ where $$E=diag(e_1,\cdots, e_n), \  F=diag(f_1,\cdots,
f_n),$$ $$E'=diag(e'_1,\cdots, e'_n), \  F'=diag(f'_1,\cdots,
f'_n),$$ and $e_i f'_i - f_i e'_i=1$, for any $i$; in addition, the rows of
$\begin{pmatrix} Id_n &\mid & N  \end{pmatrix}$ and
$\begin{pmatrix} Id_n &\mid & M  \end{pmatrix}Y$ are orthogonal.

In order to rephrase these conditions, by abuse of notation,
consider each diagonal block as a vector. That is $E=(e_1, \cdots
e_n), E'=(e'_1, \cdots e'_n)$, $F=(f_1, \cdots f_n)$ and finally
$F'=(f'_1, \cdots f'_n)$. Also for two vectors $v, u\in
\mathbb{F}^n_p$, let $v\times u$ be the vector satisfying
$$(v\times u)_i=v_iu_i,$$ and recall that $v.u$ is the usual inner
product that we used latter.

Using these notations, one can check that the orthogonality and
the determinant assumptions are equivalent to the followings;

\begin{equation}E'\cdot (M_i\times N_j)-F' \cdot (M_i\times \delta_j)+E \cdot
(\delta_i\times N_j) -F \cdot (\delta_i\times \delta_j)=0,\ \
\  \label{orth}
\end{equation}
for all $i,j$, and
\begin{equation} E\times F'-E'\times F=(1,1,\cdots, 1), \label{det1}
\end{equation}
where, $M_i$ and $N_j$ are $i$-th and $j$-th rows of matrices $M$
and $N$, respectively, and $\delta$ is the {\it Kronecker} delta
function.

(\ref{orth}) is a set of linear equations with undetermined
variables $E,E', F$ and $F'$, and its solutions consist a vector
space. Hence, one can compute a basis $\mathcal{B}$ for this
space using efficient algorithms. It means that, our problem is
reduced to checking the equation (\ref{det1}) for these
solutions. But, the space of solutions may have a large
dimension, and it may take an exponential time to check it for all
solutions. On the other hand, it is proved in \cite{us} that if
the dimension of the solutions is large enough, then, there exists
an affine subspace of {\it large} dimension satisfying equation
(\ref{det1}). In fact we have the following precise theorem (see
\cite{us}).

\begin{thm}\label{alg}{ If two graphs $G$ and $H$ are equivalent,
then there exists an affine linear subspace in the space of
solutions ( of (\ref{orth})) satisfying (\ref{det1}), with
co-dimension at most $5$.
}\end{thm}

Roughly speaking, if $G$ and $H$ are equivalent then almost all of
the solutions of (\ref{orth}) satisfy (\ref{det1}). In fact,
using the following lemma which is proved in \cite{us}, we can
check the equation (\ref{det1}) in polynomial time.

\begin{lem}\label{final} { For any base $\mathcal{B}$ of a linear space $\Lambda$, and every affine
subspace $\Gamma$ of $\Lambda$ of $codim \leq 5$, there exists a
vector $u \in \Gamma$, which is a linear combination of at most
five elements of $\mathcal{B}$.}
\end{lem}

Putting theorem \ref{alg} and lemma \ref{final} together, we
obtain the following algorithm for recognizing local equivalency
of graphs.

\vspace{5mm}

\noindent\textbf{Algorithm.} First, compute a basis $\mathcal{B}$
for the space of solutions of (\ref{orth}). Next, consider all
vectors which are a linear combination of at most $5$ vectors in
$\mathcal{B}$, and check the equation (\ref{det1}) for them. If
among them at least one satisfies (\ref{det1}), then $G$ and $H$
are equivalent, otherwise they are not equivalent. Notice that,
this is a polynomial time algorithm.


\section{Conclusion}

We have established a lower bound on the number of graph states
over $n$ qubits. For non-binary case, we have shown that the
action of local Clifford group on the graph states, can be
described by some operations on graphs. Also, we have established
an efficient algorithm to verify whether two graphs are locally
equivalent or not.

\noindent{\bf Acknowledgement.} Authors are greatly thankful to
Peter W. Shor, for all his gracious support and helpful advice.
They are also thankful to Isaac Chuang for introducing this
problem, and for all useful comments he kindly gave them.


\begin{thebibliography}{}

\bibitem{knill} A. Ashikhmin and E. Knill, {\it Nonbinary quantum stabilizer codes},
 IEEE Trans. Info. Theory, 47 (2001), 3065-3072.

\bibitem{us} M. Bahramgiri, S. Beigi, {\it An efficient algorithm to recognize locally equivalent
graphs in non-binary case}, cs/0702057.


\bibitem{bouchalg} A. Bouchet, {\it An efficient algorithm to recognize locally equivalent graphs},
Combinatorica, 11 (1991), 315-329.


\bibitem{bouchtree} A. Bouchet, {\it Transforming trees by successive local complementations},
J. Graph Theory, 12 (1988), 195-207.

\bibitem{bouchet} A. Bouchet, {\it Recognizing locally equivalent graphs},
Discrete Math., 114 (1993), 75-86.

\bibitem{cleve} R. Cleve, {\it Quantum Stabilizer Codes and Classical Linear
Codes}, Phys. Rev. A 55, 4054 - 4059 (1997).

\bibitem{erik} J. Dehaene, E. Hostens and B. De Moor, {\it Stabilizer states and Clifford
operations for systems of arbitrary dimenstions, and modular
arithmetic}, quant-ph/0408190.


\bibitem{moor2} J. Dehaene, M. Van den Nest and B. De Moor, {\it An efficient algorithm to
recognize local Clifford equivalence of graph states},
quant-ph/0405023.

\bibitem{moor1} J. Dehaene, M. Van den Nest and B. De Moor, {\it Graphical description
of the action of local Clifford trasformations on graph states},
quant-ph/0308151.

\bibitem{hein} M. Hein, J. Eisert and H. J. Briegel,
{\it Multi-party entanglement in graph states}, quant-ph/0307130.

\bibitem{marc} M. Hein, W. Dur, J. Eisert, R. Raussendorf, M. Van den Nest and H. J. Briegel,
{\it Entanglement in graph states and its applications},
quant-ph/0602096.


\bibitem{ketkar} A. Ketkar, A. Klappenecker, S. Kumar and P. K. Sarvepalli,
{\it Nonbinary stabilizer codes over finite fields},
quant-ph/0508070.

\bibitem{chuang} M. A. Nielsen and I. L. Chuang, {\it Quantum computation and
quantum information}, Cambridge University Press, (2000).

\bibitem{otter} R. Otter, {\it The number of trees},
Ann. of Math. (2), 49 (1948), 583-599.


\bibitem{sch} D. Schlingemann, {\it Stabilizer codes can be realized as graph codes},
quant-ph/0111080.

\end{thebibliography}
\end{document}